\documentclass[pra,twocolumn,showpacs,preprintnumbers,superscriptaddress]{revtex4}

\usepackage{graphicx}
\usepackage{dcolumn}
\usepackage{bm}
\usepackage{bbm}
\usepackage{psfrag}
\usepackage{multirow,amssymb,amsbsy,amsmath}
\usepackage{stmaryrd}
\usepackage{color}
\newcommand{\I}{\textup{i}}

\newcommand{\ddim}{\udelta\kern0.1em}

\newcommand{\beikonst}[2]{\left( #1 \right)_{\kern-0.2em #2}}

\newcommand{\trtxt}[2][]{\text{Tr}_{#1}\{#2\}}

\newcommand*{\ket}[1]{\mathopen{|}#1\mathclose{\rangle}}

\newcommand{\comutxt}[2]{[#1,#2]}
\newcommand{\smallotimes}{{\scriptstyle\otimes\,}}
\newcommand*{\Bra}[1]{(#1|}
\newcommand*{\Ket}[1]{|#1)}
\begin{document} 

%
%
\title{A cyclic cooling algorithm}

\author{Florian Rempp}
\email{Florian.Rempp@itp1.uni-stuttgart.de}
\affiliation{Institut f\"ur Theoretische Physik I, Universit\"at Stuttgart, %
             Pfaffenwaldring 57, 70550 Stuttgart, Germany}%
\author{Mathias Michel}
\affiliation{Advanced Technology Institute, School of Electronics and Physical Sciences, University of Surrey, Guildford GU27XH, United Kingdom}%
\author{G\"unter Mahler}
\affiliation{Institut f\"ur Theoretische Physik I, Universit\"at Stuttgart, %
             Pfaffenwaldring 57, 70550 Stuttgart, Germany}%
\received{\today}%

\begin{abstract}
We introduce a scheme to perform the cooling algorithm, first presented by Boykin et~al.\ in
2002, for an arbitrary number of times on the same set of qbits. We achieve this goal by adding an
additional SWAP-gate and a bath contact to the algorithm. This way one qbit may repeatedly be cooled without adding additional qbits to the system. By using a product Liouville space to model the bath contact we calculate the density matrix of the system after a given number of applications of the algorithm.
\end{abstract}

\maketitle

%
%

Algorithmic cooling is a method to obtain highly polarized spins in a spin system without cooling down the environment.
It may for example be used for medical magnetic resonance imaging to improve the resolution by cooling down a subset of nuclear spins of a patient without cooling of the patient himself, or for the preparation of the ground state of a quantum computer by means of the computer itself, that means that no external cooling mechanism would have to be attached to the system \cite{baugh2005,Brassard2005}.

The spin to be cooled down (a nuclear spin for example) has to couple weakly to the environment.
In addition one uses some rapid relaxing spins to transport energy out of the system.
The transportation of energy from the cooled spin to the others is achieved in a strictly non classical way by applying a quantum algorithm to the system, therefore the spins are further referred to as qbits.

Recently, Boykin et al.\ \cite{Boykin2002} have developed a quantum algorithm to cool down a single qbit with the aid of two auxiliary qbits.
Initially the system is prepared in an equilibrium state with all spins at the same inverse temperature $\beta(0)$.
(Note that we label all quantities belonging to the $n^{\text{th}}$ application of the algorithm by $(n)$, so the initial state and the temperature of the bath, introduced later on, are labeled $(0)$.)
By applying several quantum gate operations one spin is cooled down by transferring energy to the others.
The algorithm itself consists of a controlled NOT (CNOT) gate and a controlled swap gate (CSWAP) \cite{Nielsen}.
The CSWAP is a 3 qbit gate which swaps qbit 1 with qbit 3 if qbit 2 is $\ket{0}$, otherwise it does nothing.
This leads to an increase of the inverse temperature $\beta(1)$ of qbit 1 to approximately $3/2~\beta(0)$ (cf.~\cite{Boykin2002,Fernandez(2004)}).
Having applied the algorithm once, the initial state is recovered by two further applications of the algorithm.
However, by cooling down two other qbits by applying the same algorithm as described above to two additional sets of 3 qbits allows a second application of the algorithm to the cooled qbit triple with reduced initial inverse temperature $\beta(1)$.
Thus, one qbit could be cooled down to the total of $9/4~\beta(0)$.
Possessing an unlimited number of qbits the method is, in principle, able to reach arbitrary low temperatures for a single qbit, but, of course, due to an exponential growth of resources \cite{schulman2005}.

As an improvement in the same paper \cite{Boykin2002} the use of rapidly thermalising qbits is suggested.
That means that after the application of the algorithm qbit 2 and 3 are coupled to a polarization heat bath with inverse temperature $\beta(0)$ and thus relax back to their initial state (infinite bath contact time).
These two qbits can now be used to cool down a second and third qbit to use those for a second cooling step for one of the cooled ones.
This system is not closed any more and so Shannon's bound does not apply.
But even so the growth of the number of needed qbits is slowed down it is nevertheless exponential.

\begin{figure}
    \psfrag{qbit 1}{qbit 1}
    \psfrag{qbit 2}{qbit 2}
    \psfrag{qbit 3}{qbit 3}
    \psfrag{SWAP}{SWAP}
    \psfrag{CNOT}{CNOT}
    \psfrag{CSWAP}{CSWAP}
    \psfrag{Bath}{bath}
    \psfrag{b}{\footnotesize$\beta(0)$}
    \includegraphics[width=8cm]{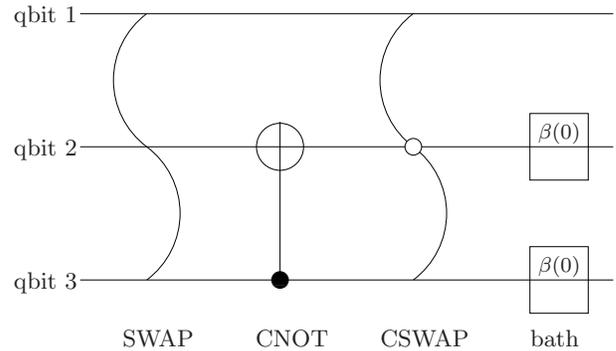}
    \caption{Cyclical cooling algorithm: 3 quantum gates are applied, first a SWAP gate then a CNOT gate and finally a CSWAP gate. Bath contact at inverse temperature $\beta(0)$ is symbolized by the boxes on qbit 2 and 3.}
    \label{algbild}
\end{figure}

Instead of using an infinite number of qbits to reach an arbitrary low temperature it would be highly desirable to have an algorithm which reaches at least some lower temperature without using more qbits.
As will be shown below, this is obtained by starting with a SWAP gate between qbit 1 and 3.
After the SWAP again a CNOT and CSWAP gate is applied to the system in the same manner as in Boykins algorithm.
Now one has to wait until the auxiliary qbits 2 and 3 are relaxed back to the initial temperature by the coupling to a heat bath (see Fig.~\ref{algbild}).
For this last step we investigate two cases: Total relaxation to the bath temperature (infinite contact time) and a finite coupling time $\tau$ to the bath, that means coherences in the system are not totally damped and the initial occupation probabilities are not entirely regained.

This new algorithm can be applied for an arbitrary number of times to the same set of qbits.
We thus call this algorithm cyclical, despite the fact that after one application the system does not return to its initial state.
We are then interested in the final inverse temperature $\beta(n)$ of the first qbit after the $n$-th application in dependence of system parameters and relaxation times between subsequent application steps.
In principle, the complete process could be seen in the context of thermodynamical machines (refrigerators) cooling down a finite ``environment'' (here only the single spin 1).

Considering more than three qbits this cyclic algorithm is able to cool down one half of the accessible qbits by applying it on sets of 4 qbits, first cooling qbit 1 with the aid of qbit 2 and 3, then cooling qbit 4 with the qbits 2 and 3, again.
If one was capable of running quantum gates on any combination of qbits, all but two qbits of the system may be cooled down.

The non-interacting 3 spin system is described by the Hamiltonian
\begin{equation}
 \hat{H}=\sum_{\mu=1}^3 \Delta E_{\mu} \hat{\sigma}_z(\mu)
\end{equation}
where $\Delta E_{\mu}$ specifies the respective Zeeman splitting of qbit $\mu$.
Each quantum gate is represented by a unitary transformation $\hat{U}$.
Thus we introduce the transformation operators $\hat{U}_{\text{SWAP}}$, $\hat{U}_{\text{CNOT}}$ and $\hat{U}_{\text{CSWAP}}$, representing the whole algorithm as $\hat{U}=\hat{U}_{\text{CSWAP}} \hat{U}_{\text{CNOT}} \hat{U}_{\text{SWAP}}$.
The density matrix after the application of the algorithm, but before the next bath contact, is given by
\begin{equation}
	\label{trafo}
	\hat{\rho}_{\text{f}}(n)=\hat{U}\hat{\rho}_{\text{i}}(n) \hat{U}^{-1},
\end{equation}
where $n$ denotes the number of applications of the algorithm until now.

Then the bath contact has to be taken into account in order to calculate the final temperature.
Since the qbits are uncoupled, each one is thermalized separately.
A standard technique to describe such a bath coupling refers to the quantum master equation in Lindblad form \cite{Breuer, Weiss, Scully, henrich(2006)}.
For a single spin $\mu$ the respective Liouville von Neumann equation reads
\begin{align}
	\label{liouvillebath}
	\dot{\hat{\rho}}_{\mu}
	=&-\I\comutxt{\hat{H}}{\hat{\rho}_{\mu}}\notag\\
	 &+W_{1\rightarrow 0}(2\hat{\sigma}_-\hat{\rho}_{\mu}\hat{\sigma}_+
	 -\hat{\rho}_{\mu}\hat{\sigma}_+\hat{\sigma}_- 
	 -\hat{\sigma}_+\hat{\sigma}_-\hat{\rho}_{\mu})\notag\\ 
         &+W_{0\rightarrow 1}(2\hat{\sigma}_+\hat{\rho}_{\mu}\hat{\sigma}_-
	 -\hat{\rho}_{\mu}\hat{\sigma}_-\hat{\sigma}_+ 
	 -\hat{\sigma}_-\hat{\sigma}_+\hat{\rho}_{\mu}),
\end{align}
according to the rates $W_{1\rightarrow 0}= \lambda/(1+(1/\varepsilon))$ and $W_{0\rightarrow 1}= \lambda/(1+\varepsilon)$, $\varepsilon = \text{exp}[\Delta E\beta(0)]$ and the bath coupling strength $\lambda$.
(Here we used for all numerical investigation $\lambda=0.01$.)

The total Liouville von Neumann equation can be represented by the superoperator $\mathfrak{L}$ acting on operators of the Hilbert space $\mathcal{H}$, here the density operator
\begin{equation}
 \dot{\hat{\rho}}_{\mu}=\mathfrak{L}\hat{\rho}_{\mu}.
\end{equation}
Sorting the entries of an operator $\hat{O}$ on $\mathcal{H}$ (for example the density operator) into a $k^2$ dimensional vector (with $k$ being the Hilbert space dimension), we define "ket" and "bra" like vectors $\hat{O} \rightarrow \Ket{\hat{O}}$ and $\hat{O}^{\dag} \rightarrow \Bra{\hat{O}}$ in this super space.
Their inner product is defined as $(\hat{A}|\hat{B})=\trtxt{\hat{A}^\dag\hat{B}}$, the trace norm of operators in $\mathcal{H}$.
Operators $\mathfrak{O}=\Ket{\hat{A}}\Bra{\hat{B}}$ acting on states $\Ket{\hat{O}}$ in the Liouville space are defined as (cf.~\cite{michel(2004)})
\begin{equation}
	\mathfrak{O}\Ket{\hat{O}}
	=\Ket{\hat{A}}(\hat{B}|\hat{O})
	=\trtxt{\hat{B}^\dag\hat{O}}\hat{A}.
\end{equation}
where the superoperator $\mathfrak{O}$ represents a $k^2 \times k^2$ dimensional matrix.
For a single spin a convenient basis is given by the Pauli operators $\hat{\sigma}_i$ with $i=\{x,~y,~z,~0\}$ ($\hat{\sigma}_0$ to represent identity).
Transferring those Hilbert space operators into the Liouville space as described above each superoperator may be expanded as
\begin{equation}
 \mathfrak{O}=\sum_{ij} O_{ij} \Ket{\hat{\sigma}_i}\Bra{\hat{\sigma}_j}.
\end{equation}
One of the big advantages of this super space formalism is the possibility of writing down a superoperator projecting an arbitrary state on a solution of Eq.~(\ref{liouvillebath}).
Just like in Hilbert space the time evolution operator [the formal solution of Eq.~(\ref{liouvillebath})] is given by
\begin{equation}
	\label{thermalisingop}
	\hat{\rho}(t)
	=\text{e}^{\mathfrak{L}t}\hat{\rho}_{\text{f}}(n)
	=\mathfrak{T}(t)\hat{\rho}_{\text{f}}(n)
\end{equation}
with the limit
\begin{equation}
 \lim_{t \to \infty} \mathfrak{T}(t)=\mathfrak{T}(\infty)
\end{equation}
defining the complete thermalisation superoperator.
Based on Eq.~(\ref{thermalisingop}) we find for diagonal density operators the time evolution operator in terms of the Pauli basis
\begin{align}
 \mathfrak{T}(\tau)=&\Ket{\hat{\sigma}_0}\Bra{\hat{\sigma}_0} + e^{-2 \tau \lambda}~ \Ket{\hat{\sigma}_z}\Bra{\hat{\sigma}_z}\nonumber\\&+\left(e^{-2\tau\lambda}-1\right)\frac{\varepsilon-1}{ \varepsilon+1}~\Ket{\hat{\sigma}_0}\Bra{\hat{\sigma}_z}.
\end{align}
This superoperator represents the thermalisation process truncated after a time step $\tau$.
In this case the bath has not jet completely thermalised the spin.
To extend the thermalising superoperator to more than one spin we use a product Liouville space with the basis $\Ket{\hat{\sigma}_i}\otimes\Ket{\hat{\sigma}_j}\otimes\Ket{\hat{\sigma}_k}=\Ket{\hat{\sigma}_{i j k}}$. In this basis the respective thermalising superoperator of qbit 2 and 3 (cf.\ Fig.~\ref{algbild}) reads
\begin{equation}
 \mathfrak{T}_{2 3}(\tau)=\mathbbm{1}\otimes\mathfrak{T}(\tau)\otimes\mathfrak{T}(\tau).
\end{equation}\par
For the superoperators of the quantum gates we use the corresponding unitary transformation (\ref{trafo}) on a general density matrix expanded in Pauli matrices, thus, obtaining the superoperator $\hat{U}\hat{\rho} \hat{U}^{-1}\rightarrow \mathfrak{U}\Ket{\rho}$ in the product basis.
We get the superoperator $\mathfrak{B}=\mathfrak{T}_{2,3}(\tau)\mathfrak{U}$ for the complete algorithm and the state of the system after an arbitrary number $n$ of cycles
\begin{equation}
	\label{rhon}
	\Ket{\hat{\rho}_{\text{f}}(n)}=\mathfrak{B}^n \Ket{\hat{\rho}_{\text{i}}(0)}.
\end{equation}
This could be achieved by finding the Jordan decomposition of $\mathfrak{B}$ and take it to its $n$-th power.
Finally, we get the density operator in the Liouville space product basis.
The result is easily transformed back to the Hilbert space by using the respective Pauli product basis in the Hilbert space $\hat{\sigma}_{i j k}=\hat{\sigma}_i\otimes\hat{\sigma}_j\otimes\hat{\sigma}_k$, by inserting the coefficients $\rho_{\text{f},ijk}(n)$ of $\Ket{\hat{\rho}_{\text{f}}(n)}$ back into the expansion
\begin{equation}
 \hat{\rho}_{\text{f}}(n)=\sum_{ijk} \rho_{\text{f},ijk}(n) \,\hat{\sigma}_{ijk}.
\end{equation}

In the case of complete relaxation of the auxiliary qbits, i.e., by taking the limit for $\tau
\rightarrow \infty$ of $\mathfrak{T}_{2 3}(\tau)$, we have been able to compute the limit $n\rightarrow
\infty$ of Eq.~(\ref{rhon}).
The asymptotic inverse temperature of the cooled qbit 1 yields
\begin{equation}
 \label{final}
 \beta(\infty)=\frac{\Delta E_2+\Delta E_3}{\Delta E_1}\, \beta(0).
\end{equation}
It just depends on the Zeeman splitting of the qbits involved in the algorithm and may thus be adjusted precisely.

\begin{figure}
    \psfrag{ 1}{1}
    \psfrag{ 1.1}{1.1}
    \psfrag{ 1.2}{1.2}
    \psfrag{ 1.3}{1.3}
    \psfrag{ 1.4}{1.4}
    \psfrag{ 1.5}{1.5}
    \psfrag{ 1.6}{1.6}
    \psfrag{ 0}{0}
    \psfrag{ 100}{100}
    \psfrag{ 200}{200}
    \psfrag{ 300}{300}
    \psfrag{b/bi}{$\frac{\beta(n)}{\beta(0)}$}
    \psfrag{n}{$n$}
    \includegraphics[width=8cm]{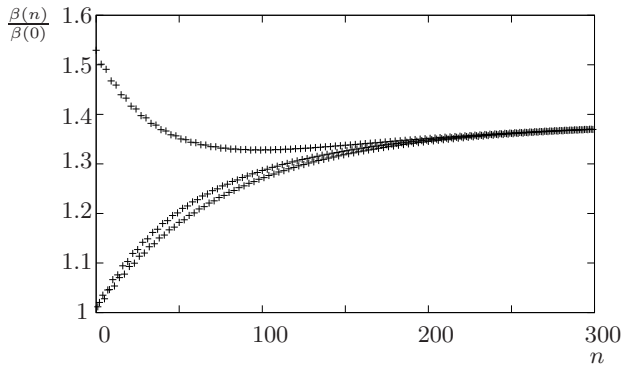}
    \caption{\label{shortbath}Trend of $\beta(n)$ as a function of cycles $n$ for short bath contact time ($\tau = 1/50\,T1$). The three curves together describe the temperature evolution of qbit 1: $\beta(n)$ jumps from the upper curve to the lowest, then to the middle one and up again.}
\end{figure}

\begin{figure}
    \psfrag{ 1.5}{1.5}
    \psfrag{ 1.6}{1.6}
    \psfrag{ 1.7}{1.7}
    \psfrag{ 1.8}{1.8}
    \psfrag{ 1.9}{1.9}
    \psfrag{ 2}{2}
    \psfrag{ 0}{0}
    \psfrag{ 10}{10}
    \psfrag{ 20}{20}
    \psfrag{b/bi}{$\frac{\beta(n)}{\beta(0)}$}
    \psfrag{n}{$n$}
    \includegraphics[width=8cm]{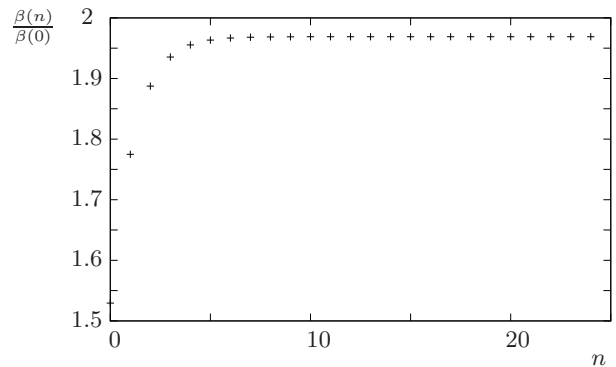}
    \caption{\label{longbath}$\beta(n)$ as a function of cycles $n$ for long bath contact time ($\tau=4\,T1$). Here the final temperature is reached approximately after a few steps.}
\end{figure}

Unfortunately, for the truncated relaxation, i.e., a finite relaxation time $\tau$, a complete analytic solution is not available, hence we investigate the algorithm numerically (with $\Delta E_1=\Delta E_2=\Delta E_3=1$).
For short bath contact time $\tau$ the inverse temperature $\beta(n)$ of the cooled spin is shown in Fig.~\ref{shortbath}, for long bath contact it evolves like Fig.~\ref{longbath}.
Apparently, in Fig.~\ref{longbath} the final stationary state is reached quickly after fewer then ten applications of the algorithm.
This is not the case for very short relaxation times, as can be seen from Fig.~\ref{shortbath}.

\begin{figure}
    \psfrag{ 1.3}{1.3}
    \psfrag{ 1.5}{1.5}
    \psfrag{ 2}{2}
    \psfrag{ 0}{0}
    \psfrag{ 100}{2}
    \psfrag{ 200}{4}
    \psfrag{b/bi}{$\frac{\beta(300)}{\beta(0)}$}
    \psfrag{t}{$\tau [T1]$}
    \psfrag{one application}{single application}
    \psfrag{perfect cooling}{perfect cooling}
    \includegraphics[width=8cm]{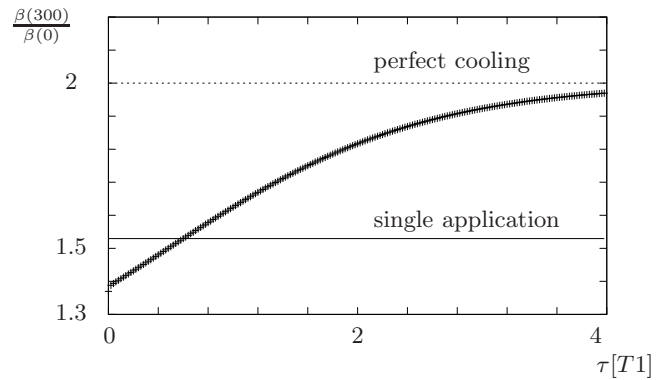}
    \caption{\label{endtemp}The final inverse temperature $\beta(300)$ as a function of different bath contact times $\tau$. The line at $\beta(300)/\beta(0)=2$ represents the upper limit for infinite bath contact, the line at $\beta(300)/\beta(0)=1,53$ marks the inverse temperature of qbit 1 after one application of the algorithm.}
\end{figure}

We measure the time in spin-latice realxation times T1.
After each application of the quantum gate we wait exactly the same time $\tau$ to let qbit 2 and 3 relax towards the equilibrium temperature of the bath.
In Fig.~\ref{endtemp} we show the final inverse temperature of qbit 1 after $n=300$ applications of the full algorithm (gates plus bath coupling) in dependence of the bath coupling time $\tau$.
Note that even if we have computed three hundred applications of the algorithm for each different waiting time $\tau$, the final temperature can already be reached after a couple of applications.
This is especially the case for $\tau\gg0.64\,T1$ as could be seen in Fig.~\ref{longbath}.

At $\tau\approx0.64\,T1$ the final temperature is as low as in the closed Boykin algorithm.
Note, that a single cycle of the new algorithm leads to the same temperature as the Boykin algorithm.
Waiting time $\tau\approx0.64\,T1$ means that qbit 2 and 3 are cooled to 0.77 times the inverse bath temperature $\beta(0)$ after the first step.

For $\tau>0.64\,T1$ the final temperature is lower than it is in the Boykin algorithm (line in Fig.~\ref{endtemp} at $\approx3/2$).
Thus, the cooling of the algorithm is improved, compared to a single cycle.
However, a multiple application of the algorithm does not always lead to a reduction of the final temperature of the cooled spin at least if $\tau<0.64\,T1$.
This can be seen from Fig.~\ref{shortbath} as well, where the temperature after the first application is already smaller than after 300 cycles.

For an easier comparison, we summarize the requirements and final temperatures of the discussed algorithms in the following table
\begin{center}%
\begin{tabular}{p{3.2cm}|l|l|l}
    Algorithm & qbits & cycles & final $\beta(n)$ \\
    \hline
    Boykin    & $3$          & $n=1$  & $\sim 3/2\beta(0)$\\
    Boykin    & $3$          & $n=2$  & $=\beta(0)$\\
    Boykin unlim.\ res.\   & $3^N$ & $n=3^N$ & $\sim (3/2)^N\beta(0)$\\
    Boykin+Bath    & $2+3^{N-1}$ & $n=3^N$ & $\sim (3/2)^N\beta(0)$\\
    \hline
    Cyclic ($\tau=$arb.)        & $3$   & $n=1$ & $\sim3/2\beta(0)$\\
    Cyclic ($\tau=\infty$)        & $3$   & $n=\infty$ & $=2\beta(0)$\\
    Cyclic ($\tau=1/50\,T1$) & $3$ & $n=300$   & $=1.37\beta(0)$\\
    Cyclic ($\tau=4\,T1$) & $3$ & $n=300$   & $=1.97\beta(0)$\\
    Cyclic ($\tau=4\,T1$) & $3$ & $n=6$   & $=1.96\beta(0)$
\end{tabular}
\end{center}

\begin{figure}
    \psfrag{ 1}{1}
    \psfrag{ 0}{0}
    \psfrag{-1}{-1}
    \psfrag{ 0.1}{0.1}
    \psfrag{ 0.5}{0.5}
    \psfrag{ 0.9}{0.9}
    \psfrag{ 1.6}{1.6}
    \psfrag{ 2}{2}
    \psfrag{ 3}{3}
    \psfrag{e}{$\eta_1$}
    \psfrag{DE2}{$\Delta E_2$}
    \psfrag{DE3}{$\Delta E_3$}
    \includegraphics[width=8cm]{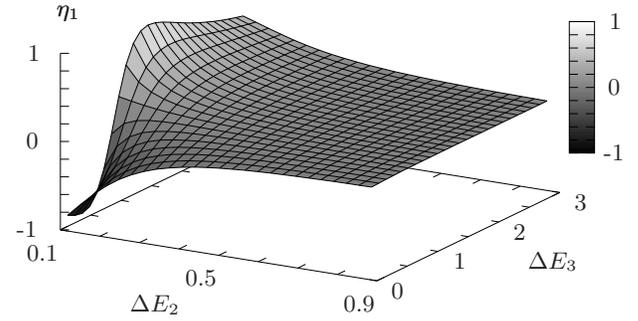}
    \caption{\label{eff}The efficiency $\eta_1$ of the first step $n=1$ of the algorithm}
\end{figure}

Finally, we calculate the efficiency $\eta$ of the algorithm. Qbit 1 represents hereby the ``second heat bath'', the most elementary bath one can think of.
The change of the energy expectation value of qbit 1 represents the heat
\begin{equation}
	\label{energieexp1}
	Q_n=\Delta\langle E(n)\rangle_1
	=\text{Tr}\{\hat{H}_1\hat{\rho}_{\text{f}}(n)\}-\text{Tr}\{\hat{H}_1\hat{\rho}_{\text{i}}(n)\}
\end{equation}
transferred by the algorithm, with $\hat{H}_1=\Delta E_1\hat{\sigma}_z\smallotimes\hat{\sigma}_0\smallotimes\hat{\sigma}_0$, $\hat{\rho}_{\text{i}}(n)$ and $\hat{\rho}_{\text{f}}(n)$ representing the density operator of the system before and after the $n$-th application of the algorithm without taking the bath contact into account.
To compute the work $W$ in the system one has to take a look at qbit 2 and 3. The energy difference is given by
\begin{equation}
	\label{energieexp23}
	\Delta\langle E(n)\rangle_{23}
	=\text{Tr}\{\hat{H}_{2 3}\hat{\rho}_{\text{f}}(n)\}
	-\text{Tr}\{\hat{H}_{2 3}\hat{\rho}_{\text{i}}(n)\}
\end{equation}
with $\hat{H}_{2 3}=\Delta E_2\hat{\sigma}_0\smallotimes\hat{\sigma}_z\smallotimes\hat{\sigma}_0+\Delta E_3 \hat{\sigma}_0\smallotimes\hat{\sigma}_0\smallotimes\hat{\sigma}_z$.
If no work was done on the system, the change of the energy expectation of this subsystem given by Eq.~(\ref{energieexp23}) would be equal to the heat (\ref{energieexp1}) with opposite sign, energy would only be moved around within the system.
Supposed that $|\Delta\langle E\rangle_{2 3}|-|\Delta\langle E\rangle_1|$ was less than zero, work would be extracted from the system, if it was larger than zero work would be done on the system.
In the case that $\Delta\langle E(n)\rangle_{2 3}$ and $\Delta\langle E(n)\rangle_1$ had different signs, $|\Delta\langle E(n)\rangle_{2 3}|-|\Delta\langle E(n)\rangle_1|$ is equivalent to the change of the energy expectation value of the entire system,
\begin{align}
	\label{energieexp}
	W_n=&\Delta\langle E(n)\rangle
	= \text{Tr}\{\hat{H}\hat{\rho}_{\text{f}}(n)\}
	- \text{Tr}\{\hat{H}\hat{\rho}_{\text{i}}(n)\}\notag\\
	=&\text{Tr}\{(\hat{H}_1+\hat{H}_{2 3}) \hat{\rho}_{\text{f}}(n)\}
	- \text{Tr}\{(\hat{H}_1+\hat{H}_{2 3}) \hat{\rho}_{\text{i}}(n)\}\notag\\
	=&\underbrace{\text{Tr}\{\hat{H}_1\hat{\rho}_{\text{f}}(n)\}
	-             \text{Tr}\{\hat{H}_1\hat{\rho}_{\text{i}}(n)\}}_{\Delta\langle E\rangle_1}\notag\\
	&+\underbrace{\text{Tr}\{\hat{H}_{2 3}\hat{\rho}_{\text{f}}(n)\}
        -             \text{Tr}\{\hat{H}_{2 3}\hat{\rho}_{\text{i}}(n)\}}_{\Delta\langle E\rangle_{2 3}}.
\end{align}
Now we can define the efficiency $\eta_n$ of the algorithm by $\eta_n=-Q_n/W_n$.
In case of complete relaxation of the auxiliary qbits we have thus obtained an analytical result for the efficiency $\eta_n$, depicted for the first step for various sets of energy splittings of qbit 2 and 3 in Fig.~\ref{eff}.
In the further applications of the algorithm for $n\rightarrow \infty$ the efficiency $\eta_n$ goes asymptotically to zero, and, of course, always stays below the accounting Carnot efficiency.
In the region, where the efficiency $\eta_1$ is negative, qbit 1 heats up instead of cooling down.

In this article we have presented a \emph{cyclic} cooling algorithm.
In comparison to the original algorithm introduced by Boykin et al.\ in \cite{Boykin2002} we added another quantum gate and a bath contact of qbit 2 and 3 using different contact times $\tau$.
The special arrangement of gates and bath contacts have made a cyclic application of the algorithm feasible.
For finite bath coupling times $\tau$ the ancilla qbits do not reach their initial temperature and also some correlations introduced by the gate operations remain within the system.
This is not the case for an infinite coupling time.
Using an adequate $\tau$ the final temperature is lower than the temperature of the original algorithm and approximately reached already after less than 10 cycles.

Furthermore, the cyclic type of algorithm allows for a comparison with standard thermodynamical machines which always use cyclic processes, e.g., the refrigerator cooling a finite reservoir by using mechanical work.
Thus, we have also computed the used work and the pumped heat as well as the efficiency of the algorithm.
Finally, a comparison to the smallest quantum thermodynamical machines \cite{henrich(2006),Henrich2007} comes into reach.
However, even those non classical algorithms do not break the Carnot limit.

A further advantage refers to the minimal amount of ressources used.
The cyclic algorithm always operates on the same three qbits.
All other improved algorithms for lower temperatures are using an exponentially growing number of ressources.
Of course, in all algorithms the final temperature is basically a function of the initial one.
In order to obtain arbitrary low temperatures a combination of different cooling mechanisms is vital.

We thank M.~Henrich, H.~Schmidt, H.~Schroeder, J.~Teifel, G.~Reuther, H.~Weimer, P.~Vidal and V.~Scarani for fruitful discussions. Financial support by the DFG is gratefully acknowledged.

\bibliographystyle{apsrev}

\end{document}